\documentclass[prl,aps,preprint,epsfig]{revtex4}
\usepackage{graphicx}% Include figure files
\usepackage{dcolumn}% Align table columns on decimal point
\usepackage{bm}% bold math
\makeatletter

\newcommand{\Rmnum}[1]{\expandafter\@slowromancap\romannumeral #1@}
\makeatother
\begin{document}
\title {Evidence of gate-tunable topological excitations in two-dimensional electron systems}
\author{R. Koushik$^1$, Matthias Baenninger$^{2,}$\footnote[1]{Present address: Department of Physics, Stanford University, Palo Alto}, Vijay Narayan$^{1,2}$, Subroto Mukerjee$^1$, Michael Pepper$^{2,}$\footnote[2]{Present address: Department of Electronic and Electrical Engineering, University College London, Torrington Place, London WC1E 7JE}, Ian Farrer$^2$, David A. Ritchie$^2$ and Arindam Ghosh$^1$ }
\vspace{1.5cm}

\address{$^1$ Department of Physics, Indian Institute of Science, Bangalore 560 012, India}
\address{$^2$ Cavendish laboratory, University of Cambridge, J. J. Thomson Avenue, Cambridge CB3 0HE, UK}

%\date{\today}
\begin{abstract}
\end{abstract}

%/PACS:  75.47.Gk,  73.40.-c,  75.70.-i

\maketitle

{\bf Topological defects are ubiquitous from solid state physics to cosmology,
where they drive phase transitions by proliferating as domain walls, monopoles or vortices. As quantum excitations, they often display 
fractional charge and anyonic statistics~\cite{Laughlin}, making them relevant to topologically protected quantum computation~\cite{Nayak}, but realizing a controlled
physical resource for topological excitations has been difficult. Here we report evidence of topological excitations during the localization transition in
strongly interacting two-dimensional electron systems (2DESs) in GaAs/AlGaAs heterostructures.
We find the electrical conductivity at low electron densities to follow a Berezinskii-Kosterlitz-Thouless (BKT)-like order-disorder
transition implying a gate-tunable proliferation of charged topological defects~\cite{Berezenskii,KT}. At low
temperatures, a weakening in the temperature dependence of conductivity was observed, and linked to the zero point fluctuations and delocalization of the defects. Our experiments also cast crucial insight on the nature of the ground state in strongly interacting 2DESs in presence of disorder}

At low temperatures ($T$), electrical conductivity $\sigma$ of a disordered
2DES decreases rapidly when the carrier density $n_s$ is reduced. The
dependence of $\sigma$ on $n_s$ often reflects the nature of the ground state
which is electrically insulating. A classical percolation
transition in inhomogenous 2DESs, for example, would cause $\sigma$ to vary as $\sim (n_s -
n_0)^{4/3}$ for $n_s > n_0$, where $n_0$ is the carrier density at the
percolation threshold~\cite{Sarma}. When disorder decreases, transition to the
insulating state occurs at much lower $n_s$, where the inter-particle Coulomb
interaction is strong, and corresponds to a large interaction parameter $r_s =
1/a_B^*\sqrt{\pi n_s}\gg 1$ ($a_B^*$ is the effective Bohr radius). The nature
of the ground state in an interacting 2DES is far from clear, although
possibilities of pinned charge-density waves (CDW), such as a Wigner crystal~\cite{Tanatar,Tsui,PD,Camjayi,Eguiluz},
charge stripes~\cite{Murthy,Koulakov}, or other phases~\cite{Kivelson}, have been suggested frequently. Physics of
2DES in this regime offers exotic
possibilities such as metallic ground states, diverging effective mass and
$g$-factors, and so on~\cite{Camjayi}.

In 2D, a CDW state with long range order does not exist at finite $T$. However,
states with quasi-long range order can exist, which can then undergo phase
transitions into a disordered states through BKT type transitions involving the
proliferation of topological defects~\cite{Berezenskii,KT,Thouless,Halperin,Young,Fertig}. Such transitions can in principle be
effected by varying $T$ or $n_s$, and the correlation length ($\xi$) in the disordered
state is a non-analytic function of the distance from the phase boundary in
density and temperature space:

\begin{equation}
\label{KT} \xi \propto \exp{\left[\frac{A}{(X - X_{KT})^{1/2}}\right ]}
\end{equation}

\noindent where $A$ is a constant, and $X = n_s$ or $T$. When $X > X_{KT}$,
topological defects unbind, resulting in finite conductivity in the
otherwise insulating pinned CDW state. The conductivity varies in accordance with
the number density $n_{ex}$ ($\sim 1/\xi^2$) of these defects, although an
explicit evidence of this has not been observed until recently~\cite{Ong}.

2DESs in modulation-doped GaAs/AlGaAs
heterostructures are excellent testing ground for many-body physics, because
it allows great control on the level of disorder. Similar to transverse
magnetic field ($H_\perp$)~\cite{PD}, disorder can also stabilize CDW state, even in the absence of
$H_\perp$, by suppressing the long wavelength fluctuations and arresting the
kinetic energy of the electrons~\cite{Eguiluz}. This increases the melting point ({\it i.e.}
$n_{KT}$ and $T_{KT}$), but excessive disorder may limit the order to very
short range, or produce a glassy state, highlighting the need for a subtle
balance. We use Si-doped GaAs/AlGaAs heterostructures, where disorder arises
from (1) Coulomb potential of ionized Si-atoms, and (2) unintentional
background doping. The former component was tuned by using devices from wafers
with different setback distances ($d$) between the dopants and the hetero-interface (See
Fig.~1a and Methods). It is important to prevent the device from pinching off ({\it i.e.} turn insulating)
at high $n_s$ which is the case in macroscopic systems with inhomogenous charge distribution~\cite{Tripathi,Hamilton}. Mesoscopic 2DESs are more suitable since they are relatively insensitive to long range
fluctuations in the conduction band~\cite{MBprb,MBprl}, and hence form the backbone of our
experiments.

\begin{figure}
\begin{center}
\includegraphics[scale=1,height=140mm]{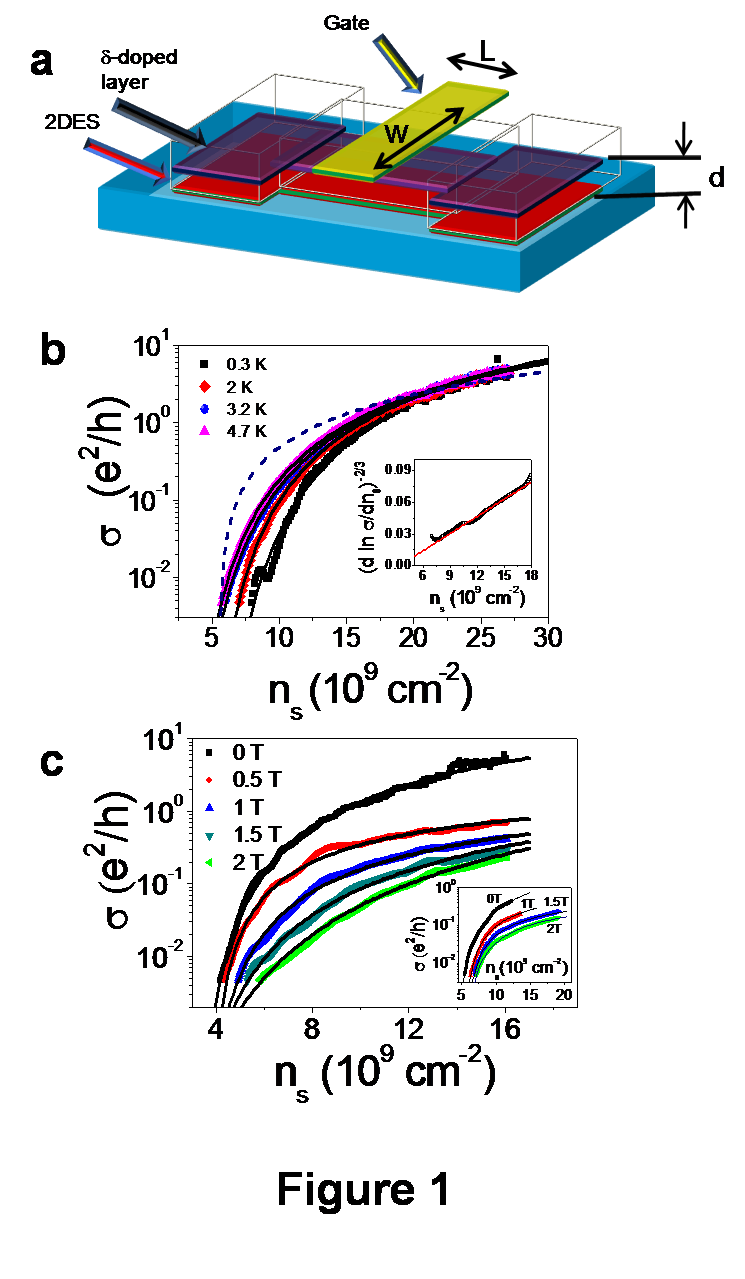}
\end{center}
\caption{\textbf{a)} Schematic of a mesoscopic device geometry showing the 2DEG (red) at the hetero-interface and the dopant layer (light blue). \textbf{b)} Condutivity $\sigma$ (device C2367) as a function of electon density ($n_s$) at different temperatures. The solid black lines are the fits for BKT transition (see Eq. $1$ in text). The dashed line represents classical percolation description. The inset shows the linearity of $(d\ln \sigma/dn_s)^{-2/3}$ in $n_s$ ($H_\perp = 0$ T, $T=3$~K) confirming the BKT analysis. The $x$-intercept gives the critical density of melting ($n_{KT}$). \textbf{c)} Conductivity $\sigma$ of device T546 vs $n_s$ at different magnetic fields and $T=262$~mK. The fit illustrates the applicability of BKT description (Eq. $1$ in text) over several decades in $\sigma$. The inset shows similar fits for another device (A2678) with different disorder. } 
 \label{figure1}
\end{figure}

The devices were patterned in the form of a microbridge with a crossed metallic
surface gate (Fig.~1a) (Ref [22]), limiting the effective area of the device to width of
$\sim 8~\mu$m and length $\sim 0.5 - 4~\mu$m. With decreasing $n_s$ (negative potential on the gate), $\sigma$
dropped rapidly in all devices, as illustrated in Fig.~1b and c. At low $T$
($\lesssim 0.5$~K), small modulations on overall smooth variation of $\sigma$
were observed occasionally, which were not reproducible on thermal cycling.
Moreover, a percolation transition also failed to describe the
$n_s$-dependence of $\sigma$ in our devices (dashed line in Fig.~1b), indicating
the influence of inhomogeneity to be small and sporadic. The pinch-off density
where $\sigma \rightarrow 0$, decreases with increasing $T$ (shown in Fig.~1b
for device C2367), but increases with increasing $H_\perp$ (shown in Fig.~1c
for T546 and A2678). The density scale at which pinch-off occurs can be very
low, $n_s \lesssim 2\times10^{9}$~cm$^{-2}$, corresponding to $r_s \simeq 13$,
and hence the possibility of a CDW ground state arises naturally.
\begin{figure}
\begin{center}
\includegraphics[scale=1,height=150mm]{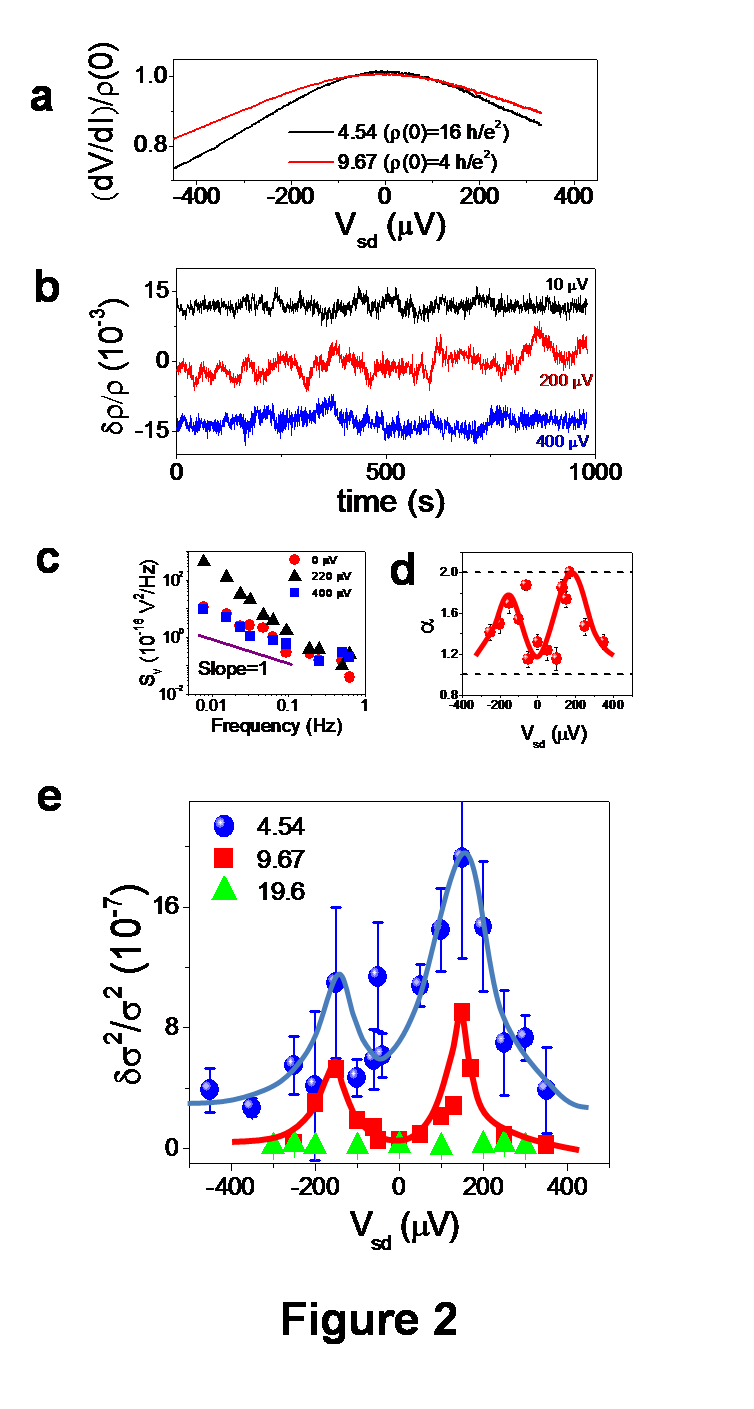}
\end{center}
\caption{\textbf{a)} Differential resistivity measurements (A2678) as a fuction of dc source-drain bias ($V_{sd}$) at $0.3$ K and two carrier densities ($10^{9}$~cm$^{-2}$) corresponding to zero-bias resistivities of $4$ $h/e^2$ and $16$ $h/e^2$. \textbf{b)} Time traces of normalized conductivity fluctuations at three different source-drain bias at $n_s = 4.54 \times 10^9$~cm$^{-2}$. Data shows significant broad band noise at bias of $V_0 \approx 200~\mu$V. \textbf{c)} The power spectral density (PSD) of conductivity noise at the three bias showing a typical $1/f^{\alpha}$ behaviour, where $\alpha$ is the spectral exponent. \textbf{d)} The plot of $\alpha$ as a function of dc bias which shows a peak magnitude of $\approx 2$ around $V_0$. The solid line is a guide to the eye. \textbf{e)} Normalized variance $\delta \sigma^{2}/ \sigma^{2}$ as a function of dc bias across the sample at different carrier densities (in units of $10^{9}$~cm$^{-2}$). The solid lines are guide to eye. The data shows a peak at around $V_0$ signifying the depinning transition of the pinned electron solid. The peak value decreases at higher densities. } 
 \label{figure2}
\end{figure}

A generic signature of pinned CDW, exploited for 1D Peierls conductors~\cite{Shobo}, as well
as magnetically stabilized 2D electron solids~\cite{Shayegan}, is the broad-band noise (BBN)
noise from the electric-field induced thermally activated transitions between the
metastable steady states. To explore this possibility, we have measured BBN in
our mesoscopic devices in a 4-probe differential (ac + dc) mode, where the
differential resistance of the mesoscopic region ($dV/dI$) was measured with a
small low-frequency ac modulation ($30~\mu$V) added onto a dc source-drain
field ($V_{sd}$). In the insulating regime (resistivity $\gg h/e^2$), $dV/dI$ decreases, but without any abrupt
threshold when $|V_{sd}|$ increases. This is illustrated for 
A2678 at $T = 262$~mK and $H_{\perp}=0$ T in Fig.~2a. However, when
$dV/dI$ was recorded as a function of time (over several hours) at fixed
$V_{sd}$, we observed a random switching noise for $|V_{sd}| \sim 100 -
200~\mu$V (see time traces in Fig.~2b). The net variance of noise ($\delta\sigma^{2}/\sigma^{2}$) hence
shows a peak at a device specific $V_{sd}$ of $V_{0} \approx 150-200~\mu$V. Away from this regime, the noise is
largely featureless with a power spectral density $\propto 1/f^\alpha$
(Fig.~2c), with the spectral exponent $\alpha \approx 1$ (Fig.~2d), arising from random potential
fluctuations at the nearby acceptor sites.  As shown in Fig.~2e, the peak in BBN decreases
rapidly with increasing $n_s$, and in the metallic regime, ($\sigma \gg e^2/h$), it decreases
below the measurement background ($\delta\sigma^2/\sigma^2 \ll 10^{-10}$). The
nonmonotonicity in noise is unlikely to arise from 
hopping, or electron glass behavior~\cite{Swastik}, but can be taken as evidence
of electrically driven depinning transition in a pinned CDW~\cite{Fertig,Shayegan}. The bias $V_0$ at
the noise maximum represents the depinning threshold which depends on the quenched
disorder, and hence only weakly dependent of $n_s$. The intermittent
plastic flow of the CDW provides the switching noise for which $\alpha$ is expected to $\approx 2$
as indeed observed (Fig.~2d).
\begin{figure}
\begin{center}
\includegraphics[scale=1,height=150mm]{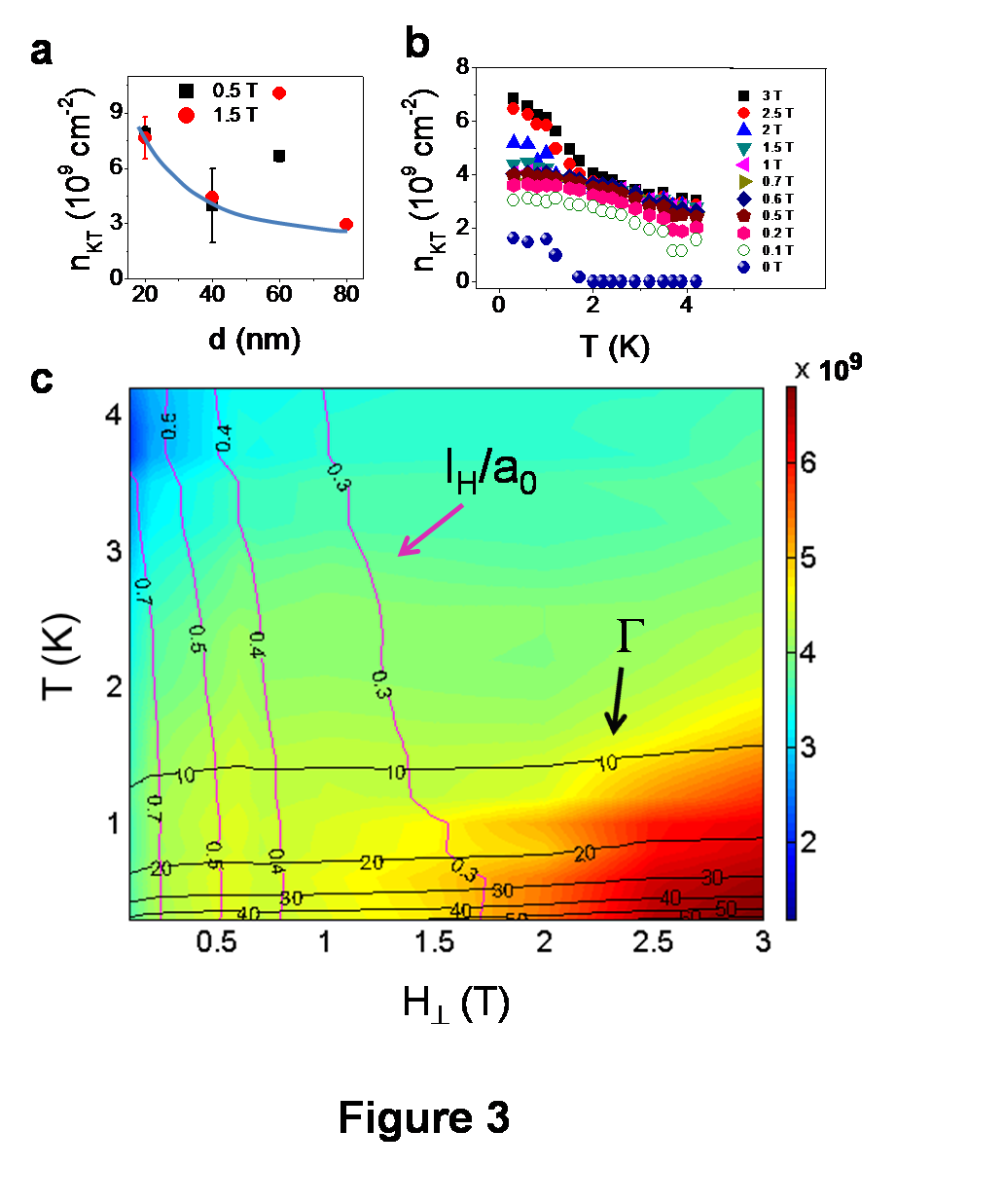}
\end{center}
\caption{\textbf{a)} Critical melting density $n_{KT}$ (at two magnetic fields) for four samples with different setback distance $d$ obtained from fitting the BKT (equation $1$ text) to conductivity. $n_{KT}$ decreases with increasing $d$, except in C2367 ($d=60$ nm) which has a strong background disorder (Methods). \textbf{b)} Evolution of $n_{KT}$ as a function of temperature at different magnetic fields ($H_\perp$). \textbf{c)} Phase diagram of $n_{KT}$ in the $T$-$H_\perp$ space. The magnitude of $n_{KT}$ is indicated by the scale bar on the right. The quantum fluctuations, characterized by the parameter $l_H / a_0$ which is the ratio of magnetic length to the inter-site distance, and the classical melting parameter $\Gamma$ (see text) are evaluated at $n_{KT}$ and some of the traces are shown. A crossover from quantum to classical melting occurs at higher magnetic fields (bottom right) where $\Gamma$ approaches $\approx 80$. } 
 \label{figure3}
\end{figure}

Evidence of topological defects was subsequently obtained by focusing on
$\sigma$ and scrutinizing its dependence on $n_s$. In accordance with Eq. 1, we find $\sigma =
\sigma_0\exp(-A/\sqrt{n_s - n_{KT}})$ describes the variation of $\sigma$ over
several decades in all devices over a wide range of $T$ (base to 4.5~K) and
$H_\perp$ (see the fits in solid line in Fig.~1b and c where $\sigma_0$ and $A$
are constants). This can be readily associated with the proliferation of unbound
charged topological defects which dominates electrical conduction in the
disordered state ($n_s > n_{KT}$). The self-consistency of the fits was also
confirmed by the linearity of $(d\ln \sigma/dn_s)^{-2/3}$ in $n_s$ (inset of
Fig.~1b), where the $x$-axis intercept gives the magnitude of $n_{KT}$ (within
10\% of that obtained from nonlinear fit).

To check if the analysis is physically meaningful, we constructed a melting
phase diagram with $n_{KT}$ extracted from the fits. Disorder quenches the
kinetic energy of interacting electrons, and stabilizes the ordered
state. Hence, decreasing the spacer thickness $d$ generally resulted in
increasing $n_{KT}$, unless the background doping dominates the disorder level.
As shown in Fig.~3a, $n_{KT}$ decreases from A2407 ($d = 20$~nm) to T546 ($d =
80$~nm), but abruptly large for C2367 ($d = 60$~nm) most likely due to strong background
doping (see Methods). In a given device though, disorder is fixed and
$n_{KT}$ depends strongly on $T$ and $H_\perp$, as illustrated for A2678 in
Fig.~3b. Note that at $H_\perp = 0$~T, $n_{KT}$ is nearly $1\times10^{9}$ cm$^{-2}$ at low
$T$, and corresponds to $r_s \approx 18$ that is much smaller than the expected
$r_s$ for Wigner crystallization in 2DESs without disorder~\cite{Tanatar}.

Fig.~3b illustrates that even a small $H_\perp$ can increase the melting point
dramatically by quenching of quantum fluctuations of the constituent ``atomic
sites''. It agrees with the recent suggestion that melting of a quantum
electron solid needs to be addressed in terms of $l_H/a_0$ rather than $n_s$
alone~\cite{Tsui}. (Here $l_H = \sqrt{h/eH_\perp}$ and $a_0 = 1/\sqrt{n_{s}}$ are the
magnetic length and mean electron distance, respectively). At higher $H_\perp$
($\gtrsim 1.5$~T), a further enhancement in $n_{KT}$ appears at low $T$,
pointing towards a different mechanism governing the stability of the CDW. To
understand this, we form a surface plot of $n_{KT}$ in $T - H_\perp$ space,
where the traces of constant classical ($\Gamma$ $=
e^2\sqrt{n_{s}}/4\pi\epsilon_0\epsilon_r k_BT$) and quantum ($l_H/a_0$) melting
parameters at $n_s = n_{KT}$ are shown (Fig.~3c). The
top-left corner, where both thermal and quantum fluctuations are strongest, the
melting occurs at small $n_{KT}$, while the bottom-right of the
plot represents maximal stability. The abrupt increase in $n_{KT}$ for $H_\perp
\gtrsim 1.5$~T occurs when $l_H/a_0 \lesssim 0.3$, signifying very little
overlap of neighboring wavefunctions and a crossover from quantum to classical
electron solid. Melting in this regime is described by $\Gamma$, which
approaches $\sim 80$ in the extreme classical limit ($H_\perp > 2.5$~T),
agreeing closely with the estimation of $\Gamma$ by Thouless for a Wigner
crystal~\cite{Thouless}.
\begin{figure}
\begin{center}
\includegraphics[scale=1,height=150mm]{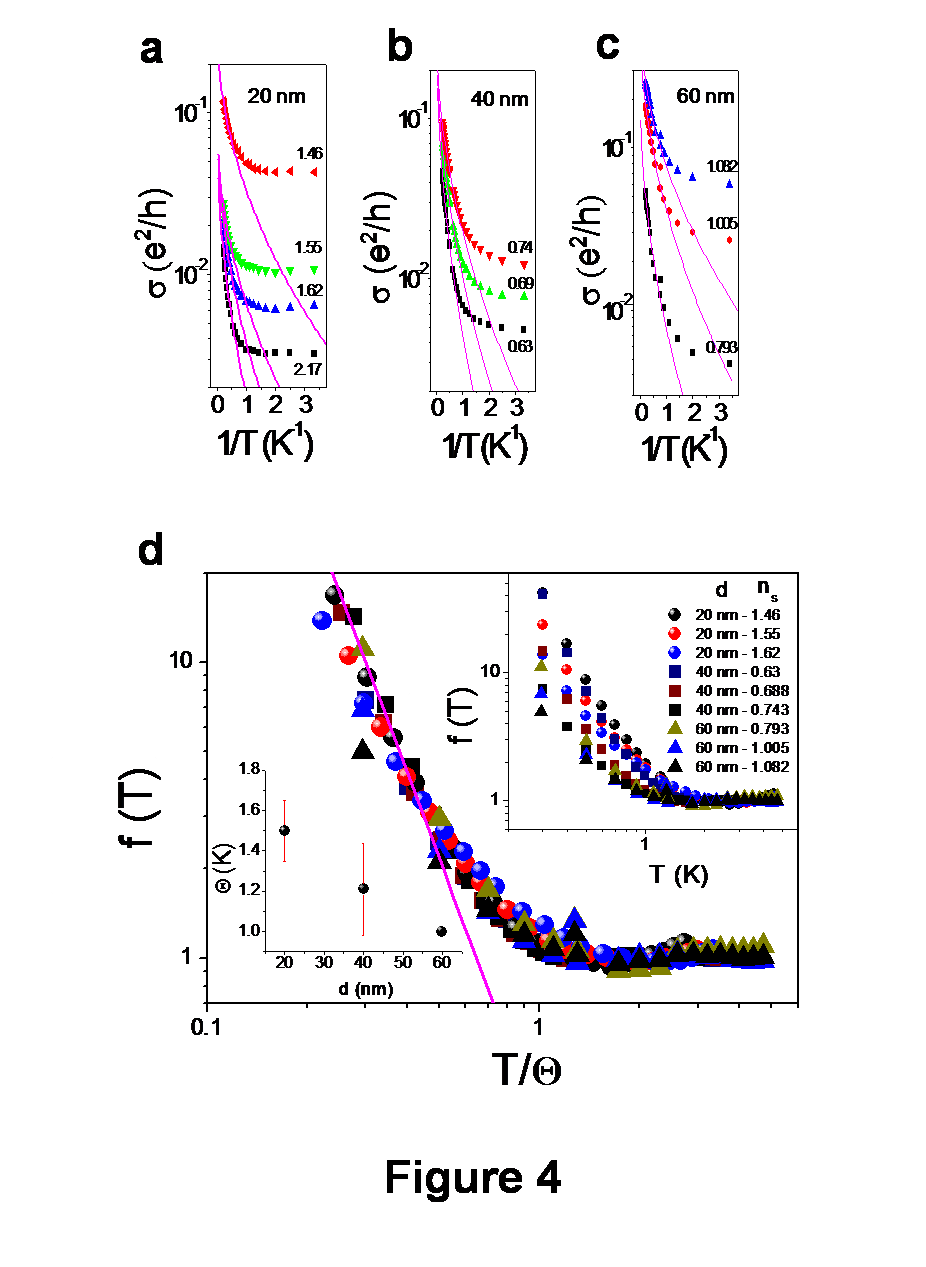}
\end{center}
\caption{\textbf{a)} Temperature dependence of conductivity $\sigma$ in three devices of different setback distances at various densities (in units of $10^{10}$ cm$^{-2}$). The solid lines are fits at the higher temperature range ($T>1.5$~K) for temperature induced BKT transition (Eq. 1). \textbf{d)} Scaling of scattering function $f(T)$ obtained by dividing $\sigma$ by the BKT fit (see text) for different devices and densities. A single parameter scaling can be observed (holding the parameter $\Theta$ for A2678 at $6.3 \times 10^9$~cm$^{-2}$ as reference). The line indicates an asymptotic behaviour of $f(T) \approx 1/T^3$ for $T \ll \Theta$. The top right inset shows experimental $f(T)$ in various devices at different $n_s$ ($10^{10}$~cm$^{-2}$). The bottom left inset shows the scaling parameter $\Theta$ to increase with increasing disorder.  } 
 \label{figure4}
\end{figure}

The $T$-dependence of $\sigma$ displays a peculiar weakening below a
characteristic temperature scale $\Theta \sim 1$~K (Fig.~4a-c) which needs further inspection. This behavior
is common to all devices, and the magnitude of $\Theta$ was found to be
insensitive to changing device length (over a factor of $\sim 8$),
eliminating finite size effect as the cause. The dependence of $\sigma$ on
$n_s$ also indicates $n_s$ to be well-defined, and an explanation based on
inhomogeneity-related models does not seem to apply here~\cite{Tripathi,Hamilton}. To understand this
quantitatively, we write $\sigma(T) \propto \exp{(-B/\sqrt{T - T_{KT}})} f(T)$. In a ``Drude-like'' scenario, for instance, $\sigma(T) \propto n_{ex}(T) \tau(T)$, where $n_{ex}(T) \sim \exp{(-B/\sqrt{T - T_{KT}})}$
and $f(T)$ would represent the temperature dependence of
mean scattering time $\tau(T)$. The behavior of $f(T)$ changes at $T \sim \Theta$, which was found to be close to the phonon gap of the electron solid through the following analysis: We obtain the
domain size $L_0$ ($\approx 430$~nm) from the depinning threshold $V_0$ ($\sim 200~\mu$V in BBN measurements for A2678 in Fig.~2)~\cite{Shayegan}, from which the phonon gap
could be calculated as $\sim \hbar c_t/k_BL_0 \approx 1.0$~K (the sound
velocity $c_t \sim 4.7 \times 10^{4}$ ms$^{-1}$ is taken to be that of a clean Wigner crystal)~\cite{Ferconi}.

The behavior of $\sigma(T)$ can be explained if $\tau(T) (\sim f(T))$ is
determined by an Andreev-Lifshitz (AL)-like mechanism of defect-mediated conduction
in a quantum solid~\cite{AL}. For $T \gg$ $\Theta$, $\tau \rightarrow$ constant, independent of $T$ due to extensive phonon scattering, and the $T$-dependence of $\sigma$ is determined by that of $n_{ex} (T)$ only. Hence to extract $f(T)$, we fitted $\sigma (T)$ by $\exp{(-B/\sqrt{T - T_{KT}})}$ in the range $T \gtrsim 1.5$~K (lines in Fig.~4a-c) and subsequently divided it by the fitted trace. The increase in $f(T)$
for $T \ll \Theta$ can be readily understood as freeze out of phonons and delocalization of defects as
quantum objects with $\tau \sim 1/T^p$. A single parameter scaling of the
$T$-axis collapses $f(T)$ in all devices onto a universal
curve, indicating $\Theta$ to be the only relevant energy scale (Fig.~4d). We find $f(T) \sim
\tau \sim 1/T^3$ at $T \ll \Theta$ (line) with $\Theta$ increasing with increasing disorder (i.e. decreasing $d$) (Fig.~4d(inset)).

Summarizing, there are four key results of this paper: First, both noise
measurements and scaling of $\sigma(n_s)$ indicate a quasi ordered ground state in
dilute 2DESs at low $T$. The precise nature of ordering is not known, Wigner
crystallization being a strong possibility, although it is clear that disorder
plays a crucial role in stabilizing such a phase. Second, the BKT scaling
provides a new framework within which the transition from strong localization
to metallic conduction can be viewed as an order-disorder melting transition.
This is in contrast to percolation~\cite{Sarma}, hopping~\cite{Keuls} or
Coulomb blockade~\cite{Ghosh} observed in many experiments over the years
(mostly in larger or more disordered 2DESs). Third, the BKT transition also
implies existence and proliferation of topological defects in 2DES in
semiconductors for the first time with tunable number density. Finally, the
scaling of the data described in Fig.~4d suggests that a mechanism like AL may
apply here and that the topological defects behave as quantum mechanical
entities capable of zero point motion and tunneling.

METHODS:

\noindent {\bf Wafer and device characteristics:}

\noindent The devices were fabricated from MBE (molecular beam epitaxy) grown
GaAs/Al$_{0.33}$Ga$_{0.66}$As heterostructures, where the two-dimensional
electron gas was formed $300$~nm below the surface. In all devices, delta
(monolayer)-doping of Si ($\sim 2.5\times10^{12}$~cm$^{-2}$) was implemented.
The strength of the disorder is partially determined by the thickness of the
spacer layer (undoped Al$_{0.33}$Ga$_{0.66}$As) situated between the
heterostructure interface and dopant layer. The typical as-grown electron
mobility was found between $0.9 - 3\times10^6$~cm$^2$/V.s at $n_s \approx
10^{11}$~cm$^{-2}$. The wafers T546 ($d = 80$~nm) and C2367 ($d = 60$~nm) were
grown in different chambers to observe the influence of background doping. The
details of various devices used in our experiment are shown in Table 1:

\begin{tabular}{|c|c|c|c|>{\centering}p{40pt}|c|}
\hline Wafer & \multicolumn{1}{>{\centering}p{40pt}|}{$n_{D}$} &
\multicolumn{1}{>{\centering}p{40pt}|}{$d$} &
\multicolumn{1}{>{\centering}p{40pt}|}{$\mu$} &
\multicolumn{1}{>{\centering}p{30pt}|}{$n_{BG}$}
\tabularnewline \hline \hline
 A2407 & 2.5 & 20 & $1.217$ & 0.5 \tabularnewline
\hline A2678 & 2.5 & 40 & $1.8$ & 0.5 \tabularnewline \hline C2367 & 0.7
& 60 & $1.2$ & 1.0 \tabularnewline \hline T546 & 1.9 & 80 & 0.9 & 0.3 \tabularnewline \hline
\end{tabular}\\
\newline \textbf{Table 1.} Details of devices used in our measurements showing
the doping concentration $n_{D}$ ($10^{12}$~cm$^{-2}$), spacer thickness $d$
(nm), as-grown mobility $\mu$ ($10^{6}$~cm$^{2}$/V.s) and background doping
estimate $n_{BG}$ ($10^{14}$~cm$^{-3}$).

\noindent {\bf Measurements:}

\noindent The noise measurements were carried out inside a He-3 cryostat (base
electron temperature \ensuremath{\approx}270~mK), using a four-probe ac+dc
technique with the ac bias fixed at $\approx 30~\mu$V at an excitation
frequency of 128 Hz, and dc bias varied over $\pm 400~\mu$V. The voltage
fluctuations across the sample were measured using a lock-in amplifier whose
output was subsequently digitized using a 16-bit digitizer, followed by
decimation and power spectral density (PSD) calculations using Welch's method
of averaged periodogram.

\end{document}